\documentclass[a4paper,11pt,nontitlepage]{article}
\usepackage{jheppub} 
\pdfoutput=1

\RequirePackage[numbers,sort&compress]{natbib}

\RequirePackage[utf8]{inputenc} 			
\RequirePackage[T1]{fontenc} 				
\RequirePackage[autostyle=true]{csquotes}	
\RequirePackage{graphicx}					
\RequirePackage{listings}					
\RequirePackage{xcolor}					    
\RequirePackage{siunitx}   					
\RequirePackage[english]{babel} 
\RequirePackage{hyperref}					
\RequirePackage[all]{hypcap}				
\RequirePackage{amsmath}
\RequirePackage{physics}					
\RequirePackage[varg]{txfonts}       		
\RequirePackage{caption}					
\RequirePackage{subcaption}
\usepackage{mathtools}
\RequirePackage{pgf}
\RequirePackage{booktabs}

\title{Improved  Constraints on the Variation of the Weak Scale from Big Bang Nucleosynthesis}

\author[a]{Helen Meyer}
\author[a,b]{and Ulf-G. Meißner}

\affiliation[a]{Helmholtz-Institut f\"{u}r Strahlen- und Kernphysik and Bethe Center for Theoretical Physics,\\ Universit\"{a}t Bonn, D-53115 Bonn, Germany}
\affiliation[b]{Institute for Advanced Simulation, Forschungszentrum J\"ulich, D-52425 J\"ulich, Germany}

\emailAdd{hmeyer@hiskp.uni-bonn.de}
\emailAdd{meissner@hiskp.uni-bonn.de}

\abstract{
We present an improved calculation of the  light element abundances in the framework of Big Bang
nucleosynthesis as a function of the Higgs vacuum expectation value $v$. We compare the methods of our 
calculation to previous literature including the recently published work of Burns
et al.~\cite{burns2024constraints}. 
The PDG result for the $^4$He abundance can be explained within $2\sigma$ by $-0.014 \leq \delta v / v \leq 0.026$, for deuterium we find the constraint $-0.005 \leq \delta v / v \leq -0.001$. These bounds are more stringent than what was found earlier.}

\begin{document}

\maketitle	


\section{Introduction}
Big Bang nucleosynthesis (BBN) is a fine laboratory for the possible variations of the fundamental constants of nature, see e.g.
\cite{Hogan_2000,Uzan_2003,Schellekens_2013,Meissner_2015,Donoghue_2016,Adams_2019}.
In a recent publication \cite{burns2024constraints}, Burns et al.\,studied possible constraints on a variation of the Higgs vacuum expectation value (VEV) from BBN simulations. They use the new \texttt{PRyMordial} code \cite{burns2023prymordial} to compute the Helium-4 ($^4$He) and deuterium ($d$) abundances for different values of the Higgs VEV and compare their results to the newest observation by the EMPRESS collaboration \cite{Matsumoto_2022} that differs from the values given by the Particle Data Group (PDG) \cite{PDG_Obs} and from current theoretical predictions. The possible interval for the Higgs VEV that they found reproduces the EMPRESS $^4$He abundance would, however, worsen the deviation for deuterium significantly.  This work triggered the investigations here, which improve the methodology
of earlier studies, such as \cite{Dixit:1987at, Scherrer:1992na, Yoo:2002vw, PhysRevD.73.023509, Dent:2007zu,Bedaque:2010hr,Berengut:2013nh},  and thus gives more reliable bounds on the possible variations of the weak scale. Here, we mostly consider the element abundances from the PDG as the EMPRESS data are presently not included in the PDG listings. 

Changing the Higgs VEV $v$ influences all elementary particle masses as they scale linearly with $v$, assuming that the Yukawa couplings remain unchanged.  This influences not only the neutron-proton mass difference, which is particularly significant for the $^4$He abundance,  but also the QCD scale $\Lambda_\mathrm{QCD}$. Burns et al. take the latter into account~\cite{burns2024constraints}. They also use some general scaling arguments for the Higgs VEV dependence of the pertinent nuclear interactions.  While this is certainly
a legitimate procedure, we  believe that the calculation can be improved in some parts, as shown here. In particular, the treatment of the neutron-proton mass difference, see Sect.~\ref{sec:mnp} is updated, and the quark mass dependence of the deuteron binding energy, see Sec.~\ref{sec:deut}, will be the subject of our work. In the most sophisticaed treatments, this has  been considered using chiral effective field theories either in a pionless \cite{Bedaque:2010hr} or pionfull \cite{Berengut:2013nh} approach. However, there is a sizeable uncertainty related to the quark mass variation of the four-nucleon contact interactions, which have so far been estimated using naive dimensional analysis \cite{Beane:2002vs,Epelbaum:2002gb} and also the so-called resonance saturation approach \cite{Epelbaum:2001fm,Berengut:2013nh}. Here, we overcome these limitations by combining lattice QCD data with the low-energy theorems
for nucleon-nucleon scattering~\cite{Baru:2015ira,Baru:2016evv} (and references therein). 
The weak $n\leftrightarrow p$ rates are discussed
in Sec.~\ref{sec:rates} and the $n+p\to d+\gamma$ reaction is considered in Sec.~\ref{sec:npdg}.
Our results are displayed and discussed in Sec.~\ref{sec:res}. Appendix~\ref{sec:A} contains some details 
on a one-boson-exchange model that can be used to study the deuteron for a varying Higgs VEV. 
In Appendix~\ref{sec:B}, we compare in detail the differences between our work and previous publications on this topic.

\section{Neutron-proton mass difference}
\label{sec:mnp}
The quark masses change linearly with the Higgs VEV and thus also the mass difference between up- and down-quark scales with the relative difference $\delta v / v$.
Throughout this work, we assume that the corresponding Yukawa couplings are constant.
The neutron-proton mass difference
 has a QCD part coming from the quark mass difference and a QED part coming from Coulomb effects between the charged quarks in the nucleon. To a good approximation, the latter should stay constant under variation of the Higgs VEV. The QCD contribution to the $n-p$ mass difference is not, however, equal to just the current quark mass difference \footnote{We remind the reader that the light quark masses are not RG invariant and thus not observable.} but has also contributions from the matrix element $\mel{p}{u\Bar{u} - d\Bar{d}}{p}$ \cite{Gasser:1982ap}. 
 This can be understood easily from the two-flavor QCD Hamiltonian, where the explicit
 chiral symmetry breaking originates from the mass term $m_u \bar{u}u + m_d \bar{d}d$, and the neutron-proton mass difference is given by its isovector $I=1$ component,
\begin{equation}
 {\cal H}_{\rm QCD}^{\rm mass} = m_u \,\bar u u + m_d  \,\bar  d d 
 = \underbrace{\frac{1}{2} (m_u+m_d)(\bar u u + \bar d d)}_{I = 0}
              + \underbrace{\frac{1}{2} (m_u-m_d)(\bar u u - \bar d d)}_{I = 1}~.   
\end{equation}
 In \cite{QnAlpha}, the electromagnetic contribution to the neutron-proton mass difference was precisely determined using the Cottingham sum rule and the strong part can then
 be deduced from the physical mass splitting, leading to
 \begin{equation}
     m_\mathrm{QCD} = 1.87 \mp 0.16~{\rm MeV}~, \quad m_\mathrm{QED} = -0.58 \pm 0.16~{\rm MeV}~.
 \end{equation}
 The neutron-proton mass difference as a function of the Higgs VEV should therefore be taken as
 \begin{equation}\label{eq:mnp}
     \frac{Q_N}{\si{\mega\electronvolt}} = \frac{m_n - m_p}{\si{\mega\electronvolt}} = (1.87 \mp 0.16) \left( 1 + \frac{\delta v}{v}\right) - (0.58 \pm 0.16).
 \end{equation}
 We note in passing that the separation of the strong from the electromagnetic contribution is afflicted with some problems, for a 
 pedagogical discussion see~\cite{Meissner:2022cbi}.

\section{Pion mass dependence of the deuteron binding energy and the nucleon-nucleon scattering lengths}
\label{sec:deut}

\begin{figure}
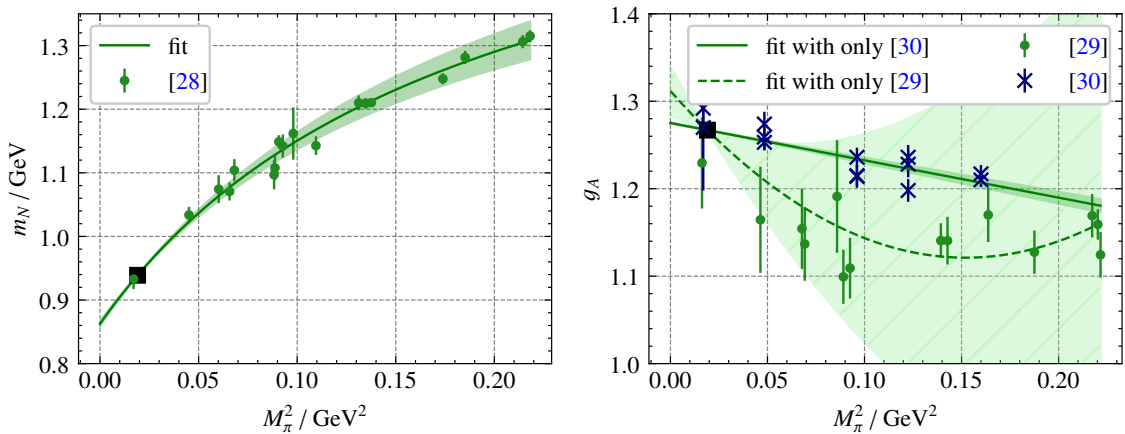

    \centering
    \resizebox{0.49\textwidth}{!}{\input{figs/mNMpi.pgf}}
    \resizebox{0.49\textwidth}{!}{\input{figs/gAMpi.pgf}}
    \caption{Rational function fit to lattice QCD data for the nucleon mass $m_N$ (\cite{Alexandrou:2013jsa}) and the axial vector coupling $g_A$ (\cite{Alexandrou:2014wca, Chang:2018uxx}) as a function of the pion mass squared. We forced the fits to go through the physical point (black filled square). The bands correspond to $1\sigma$-errors of the fit. The dashed line and hatched band display the fit for $g_A$ using only data from \cite{Alexandrou:2014wca}.}
    \label{fig:Mpifits}
\end{figure}

Changing the quark masses influences, of course, also the pion mass. To first order in chiral perturbation theory the average pion mass can be found from the Gell-Mann--Oakes--Renner relation
\begin{equation}
    M_\pi^2 = B_0 (m_u + m_d) +{\cal O}((m_u\pm m_d)^2)~, 
\end{equation}
where the quark masses scale linearly with $\delta v / v$ and the constant $B_0$, which is related to the scalar singlet quark condensate, is taken to depend on $\delta v$ like the QCD scale $\Lambda_\mathrm{QCD}$, which \cite{burns2024constraints} found to be proportional to $(1 + \delta v / v)^{0.25}$. The pion mass at a given value of the Higgs VEV is therefore
\begin{equation}\label{eq:Mpi_VEV}
    M_\pi = M_\pi^\mathrm{phys} \left( 1 + \frac{\delta v}{v}\right)^{1.25 / 2}.
\end{equation}
Here, $M_\pi^\mathrm{phys} = \SI{138.03}{\mega\electronvolt}$ is the physical average pion mass for the present value of $v$ (meaning $\delta v = 0$). For the Higgs VEV varying by $\pm 10\%$, that is $|\delta v/v| \leq 0.1$, the pion mass varies between $\SI{129}{\mega\electronvolt}$ and $\SI{147}{\mega\electronvolt}$.
A change in the pion mass affects also other physical quantities that are very relevant for BBN simulations -- above all the average nucleon mass $m_N$. The axial vector coupling $g_A$ as well as the pion decay constant (which will become relevant later) do also depend on $M_\pi$. One could use chiral perturbation theory for calculating the pion mass dependence of these parameters, as was done, e.g., in \cite{Berengut:2013nh}. However, for already moderate  changes in the pion mass, chiral perturbation theory becomes inaccurate for certain observables
in the one-baryon sector. More precisely, most lattice data can be fit well with
a linear function in $M_\pi$, see e.g. \cite{Walker-Loud:2014iea}, and the pion mass dependence of $g_A$ exhibits strong cancellations of one- and two-loop contributions~\cite{Bernard:2006te}. We therefore fit a rational function to lattice QCD data for these quantities at different values of $M_\pi$.
Quite differently, the pion mass dependence at one-loop of the pion decay constant matches the lattice data for not too large quark masses. 
Consider first the nucleon mass $m_N$. We fit the data from Ref.~\cite{Alexandrou:2013jsa} with a rational function
    \begin{equation}\label{eq:mNMpi}
        m_N(M_\pi^2) = m_N^\mathrm{phys} + \frac{\SI{3.63\pm 0.19}{\per\giga\electronvolt} \left(M_\pi^2-(M_\pi^\mathrm{phys})^2\right)}{1 + \SI{4.80\pm 0.61}{\per\giga\electronvolt\squared} \left(M_\pi^2-(M_\pi^\mathrm{phys})^2\right)}~,
    \end{equation}
 where $m_N^\mathrm{phys} = \SI{938.94}{\mega\electronvolt}$ is the value of the physical nucleon mass, that is for $\delta v =0$.  We note, however, that in the range for the Higgs VEV variation discussed here, the third order
 chiral perturbation theory (ChPT) representation of $m_N$ gives a pion mass variation consistent with the one from fitting to the lattice data. Still, we prefer to work with this fit function as it  allows for larger variations in $M_\pi$
 and thus $\delta v$.
 Similarly, for the axial-vector coupling, the data from \cite{Alexandrou:2014wca} are fitted with a polynomial 
    \begin{equation}
        g_A(M_\pi^2) = g_A^\mathrm{phys} - \SI{2.19\pm 0.27}{\per\giga\electronvolt} (M_\pi^2-(M_\pi^\mathrm{phys})^2) + \SI{8.23\pm 1.47}{\per\cubic\giga\electronvolt} (M_\pi^2-(M_\pi^\mathrm{phys})^2)^2.
    \end{equation}
Including only the more recent and more precise lattice QCD data from \cite{Chang:2018uxx}, one arrives instead at
    \begin{equation}\label{eq:gAMpi}
        g_A(M_\pi^2) = g_A^\mathrm{phys} - \SI{0.43 \pm 0.04}{\per\giga\electronvolt} \left(M_\pi^2-(M_\pi^\mathrm{phys})^2\right),
    \end{equation}
where $g_A^\mathrm{phys} = 1.267$ denotes the physical value at $\delta v=0$. 
In all these equations, $M_\pi$ is given in $\si{\giga\electronvolt}$. The fit results for $m_N$ (left panel) and $g_A$ (right panel) are displayed together with the used lattice data in Fig.\,\ref{fig:Mpifits}. The bands correspond to $1\sigma$-errors of the fit. Comparing both fits for the axial vector coupling, it is apparent that there is still quite some uncertainty when it comes to calculations on the lattice. We choose to work with the latter fit for $g_A$ that includes the more recent data. Choosing one parametrization or the other has, however, very little effect on the following calculations, because the range of pion mass variations is much smaller than what is shown in the figure.
Note also that the one-loop  ${\mathcal O}(p^3)$ ChPT result strongly deviates from the fit already in the small range of pion mass variations considered here.

The axial-vector coupling enters  as a parameter when computing the weak neutron-proton interaction rates (see below). A change in the average nucleon mass has, of course, wide-ranging effects in the BBN simulation. We change the neutron and proton mass so that they fulfill both the new value for the mass difference and the average mass:
\begin{align}
    m_p(\delta v) &= m_N(\delta v) -  Q_N(\delta v)/2 ~,\\
    m_n(\delta v) &= m_p(\delta v) + Q_N(\delta v)~.
\end{align}

In \cite{burns2024constraints}, the deuteron binding energy was computed using a  one-boson exchange model taken from \cite{MEISSNER1988213}.  
More precisely, these authors used three Yukawa functions corresponding to the pion-, omega- and sigma-exchange and essentially tuned the $\sigma$-mass
to get the corrrect deuteron binding energy at $\delta v=0$. The pion mass was scaled with the varying Higgs VEV as given above and the other meson masses as 
$(\delta v / v)^{0.25}$. It should be noted, however, that the quark mass dependence
of the $\sigma$ is more intricate than the one of a genuine quark-antiquark meson as worked out in detail in~\cite{Hanhart:2008mx}.
In the Appendix, we discuss a somewhat more sophisticated one-boson-exchange model  to better understand the approximations made in~\cite{burns2024constraints}.

\begin{figure}
	\centering
	\resizebox{0.65\textwidth}{!}{\input{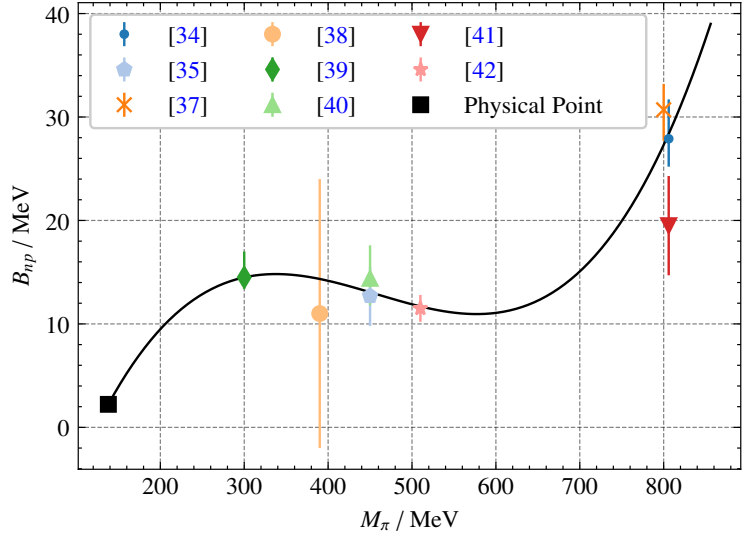}}
	\caption{Cubic polynomial fit (black line) to lattice QCD data for the deuteron binding energy in the  ${}^3S_1$ channel of nucleon-nucleon scattering. The fit was forced to go through the physical point.}
	\label{fig:BE_fits}
\end{figure}

\begin{figure}
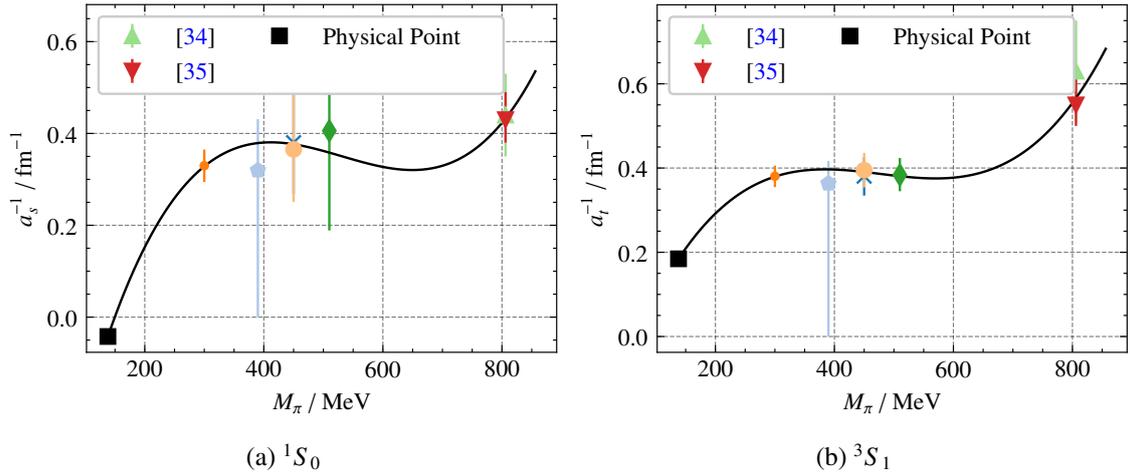

	\centering
	\begin{subfigure}{.49\textwidth}
		\resizebox{\textwidth}{!}{\input{figs/a_1S0.pgf}}
		\caption{${}^1S_0$}
	\end{subfigure}
	\begin{subfigure}{.49\textwidth}
		\resizebox{\textwidth}{!}{\input{figs/a_3S1.pgf}}
		\caption{${}^3S_1$}
	\end{subfigure}
	
	\caption{Cubic polynomial fit (black line) to inverse scattering length resulting from the LET calculations in the ${}^1S_0$ (left panel) and the ${}^3S_1$ channel (right panel) of nucleon-nucleon scattering. The data points from Refs.~\cite{Wagman:2017tmp} and \cite{NPLQCD:2020lxg} are direct lattice QCD results. The fit was forced to go through the physical point.}
	\label{fig:ainv_fits}
\end{figure}

Here, we  use lattice QCD results for the dependence of the deuteron binding energy on the pion mass. In Ref.~\cite{Lahde:2019yvr},   five data points for the $nn$ and deuteron binding energy or scattering length were compiled and the  so-called low-energy theorems (LETs), see e.g., \cite{Baru:2015ira, Baru:2016evv}, were used to calculate the scattering length at a given pion mass from the  binding energy obtained from lattice data at different quark masses without having information on the corresponding effective range.  Note that at these pion masses, the $nn$ system is also bound.
Using this method and adding two recent lattice QCD data points from \cite{Wagman:2017tmp, NPLQCD:2020lxg}, we get the pion mass dependence of the nucleon-nucleon scattering lengths  in the $S$-wave channels and the deuteron binding energy by fitting a cubic polynomial to the obtained data. The corresponding results can be found in Figs.~\ref{fig:BE_fits} and \ref{fig:ainv_fits}. The scattering lengths become relevant later for the $n + p \to d +\gamma$ reaction.  In more detail, the deuteron binding energy as a function of $x = M_\pi - M_\pi^\mathrm{phys}$ is given by
    \begin{equation}\label{eq:Bd_LET}
        B_d (M_\pi) = B_d(M_\pi^\mathrm{phys}) + \num{0.149} x - \SI{5.414 e-4}{\per\mega\electronvolt} x^2 + \SI{5.651 e-7}{\per\mega\electronvolt\squared} x^3~,
    \end{equation}
where all masses are given in units of MeV. We note that the situation about the deuteron being bound is still a disputed issue
in lattice QCD, see e.g~\cite{Amarasinghe:2021lqa}, thus we forced the fit to go to the physical value for the physical pion mass.    Further,   
the inverse scattering length in the ${}^1 S_0$ channel is
    \begin{equation}
        a_s^{-1} (M_\pi) = a_s^{-1}(M_\pi^\mathrm{phys}) + \SI{3.762 e-3}{\per\mega\electronvolt} x - \SI{1.055 e-5}{\per\mega\electronvolt\squared} x^2 + \SI{8.958 e-9}{\per\mega\electronvolt\cubed} x^3~,
    \end{equation}
while the inverse scattering length in the ${}^3S_1$ channel (corresponding to the deuteron) is given by
    \begin{eqnarray}
    a_t^{-1} (M_\pi) &=& a_t^{-1}(M_\pi^\mathrm{phys}) + \SI{2.129 e-3}{\per\mega\electronvolt\per\femto\m} x \nonumber\\ &-& \SI{6.789 e-6}{\per\mega\electronvolt\squared\per\femto\m} x^2 + \SI{6.672 e-9}{\per\mega\electronvolt\cubed\per\femto\m} x^3~.
    \end{eqnarray}

\begin{figure}
    \centering
    \includegraphics{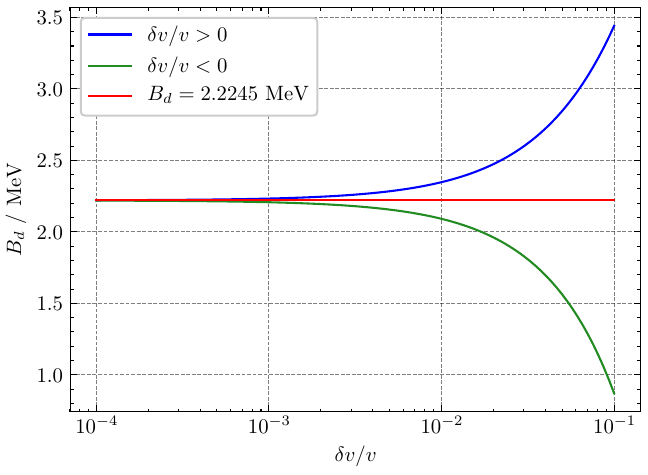}
    \caption{Deuteron binding energy as a function of the Higgs VEV, calculated using Eqs.~(\ref{eq:Mpi_VEV},\ref{eq:Bd_LET}).}
    \label{fig:Bd_LET}
\end{figure}

Calculating the deuteron binding energy in this way for different values of the Higgs VEV yields a similar result to what was found in \cite{burns2024constraints} as can be seen from comparing Fig.~\ref{fig:Bd_LET} to Fig.~2 of~\cite{burns2024constraints}. It is unclear to us why these results are so similar. Using a more sophisticated one-boson-exchange model, we indeed find a different result as discussed in appendix \ref{sec:A}.

\section{Weak $n \leftrightarrow p$ rates}\label{s:weakrates}
\label{sec:rates}

Before the universe is cool enough for deuterium fusion to be efficient and for BBN to actually start, the weak interactions between neutrons and protons are dominant. Because especially the $^4$He abundance is very sensitive to the number of neutrons existing at the beginning of BBN, it is quite important to accurately calculate the neutron-proton interaction rates.  Here, one has to consider effects from a changing electron mass  (as noted before, $m_e$ also scales linearly with $\delta v$) on the thermal background as well as the change in the $W$-boson mass that modifies the Fermi constant $G_F$:
\begin{equation}
    G_F = \frac{\sqrt{2}}{8} \frac{g^2}{M_W^2} = \frac{1}{\sqrt{2} v^2} \to G_F (\delta v) = {G_F(0)}{\left(1+\frac{\delta v}{v}\right)^{-2}}.
\end{equation}

The relevant Higgs VEV dependent part of the neutron lifetime is given by 
\begin{equation}\label{eq:tau_n}
    \tau_n \sim \frac{1}{\mathcal{G} f_R}\qc \mathcal{G} = \frac{(G_F V_\mathrm{ud})^2 (1+3g_A^2)}{2\pi^3} m_e^5,
\end{equation}
where $f_R$ is the integral over the Fermi energy spectrum. For small temperatures one can ignore temperature-dependent effects from the neutrino and electron Fermi distribution functions when calculating the neutron $\beta$-decay rate. So, in this approximation, $f_R$ is given by \cite{Segre-gamow}
\begin{align}\label{eq:Fermi-int}
    f_R &= \frac{1}{m_e^5} \int_0^{p_{e,\mathrm{max}}} \left( Q_n - \sqrt{m_e^2 + p_e^2}\right)^2 F(Z,p_e) p_e^2 \dd{p_e}, \\
    &= \int_0^{x_\mathrm{max}} \left(\sqrt{1+x_0^2} - \sqrt{1 + x^2} \right)^2 x^2 \dd{x},\,\mathrm{where} \quad x = \frac{p_e}{m_e}\quad\mathrm{and}\quad x_0 = \sqrt{\left(\frac{Q_n}{m_e}\right)^2 -1}.
\end{align}
$F(Z,p_e)$ is the Fermi function including Coulomb effects between the charged electron or positron and the proton in the final state. Of course, one could also include QED radiative corrections, finite-mass, electroweak magnetism and finite-temperature effects, all of which are considered in the calculation of the weak $n \leftrightarrow p$ rates in \texttt{PRyMordial} \cite{burns2023prymordial}. Because these are higher order effects, we will not include them in the Higgs VEV correction to the neutron lifetime.
At high temperatures the Fermi functions of the neutrino and the electron cannot be ignored, thus one has to recompute the weak neutron-proton rates every time one changes the relevant parameters (see also~\cite{Meissner:2023voo} for a detailed discussion). 
Because the neutron-proton mass difference as well as the electron mass changes with $\delta v$, the variation in $f_R$ influences the neutron lifetime significantly.
As mentioned above, the axial-vector coupling depends on the pion mass and therefore also on the Higgs VEV which should be included in the calculation. 

\begin{figure}
    \centering
    \includegraphics{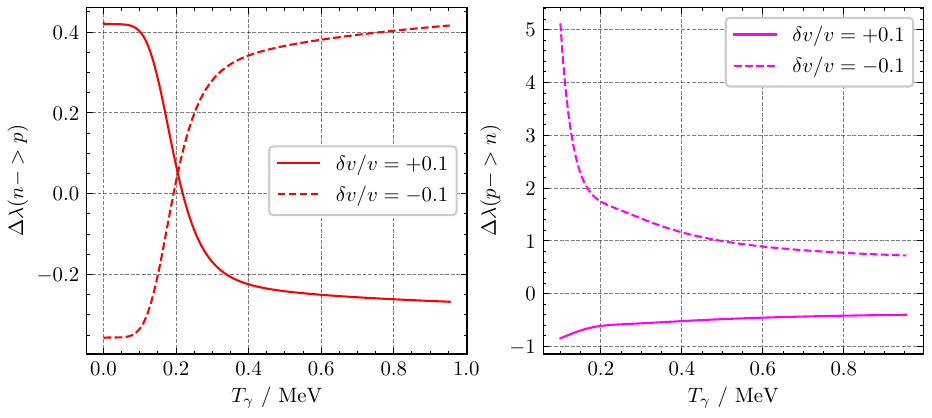}
    \caption{Fractional change in the weak $n \leftrightarrow p$ reaction rates including all relevant effects. }
    \label{fig:nTOp}
\end{figure}

\begin{figure}
    \centering
    \includegraphics{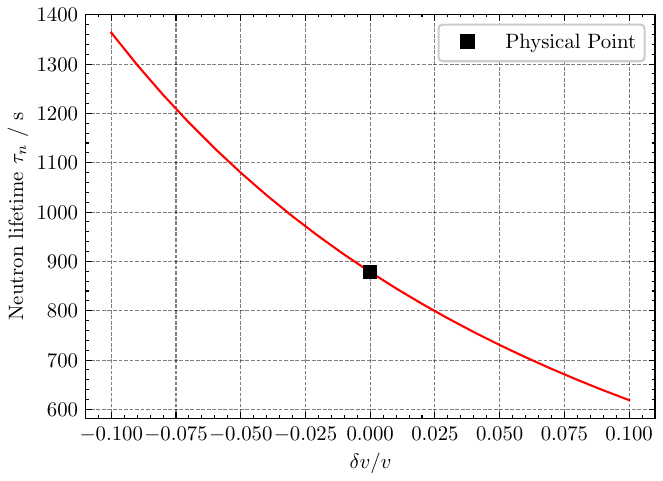}
    \caption{The neutron lifetime $\tau_n$ as a function of the Higgs VEV variation $\delta v / v$. The black square denotes the physical point $\tau_n(\delta v = 0) = \SI{878.4}{\s}$ \cite{PDG_Obs}.}
    \label{fig:tau_n}
\end{figure}

We therefore recalculate the $n\leftrightarrow p$ weak rates for every value of $\delta v / v$ changing the electron, neutron and proton mass as mentioned above as well as $G_F$ and $g_A$. The neutron lifetime changes according to
\begin{equation}
    \tau_n(\delta v) = \tau_n(0) \frac{\mathcal{G}(0) f_R(0)}{\mathcal{G}(\delta v) f_R(\delta v)}.
\end{equation}
In Fig.~\ref{fig:nTOp} the fractional change in the properly normalized $n \leftrightarrow p$ rates obtained from our calculation is displayed. Fig.~\ref{fig:tau_n} shows the neutron lifetime as a function of the Higgs VEV. We observe a significant variation from the experimental value at $\delta v = 0$, which is $\tau_n = \SI{878.4}{\s}$ \cite{PDG_Obs}.

\section{The $n + p \to d + \gamma$ reaction}
\label{sec:npdg}

\begin{figure}
    \centering
    \includegraphics[width = 0.8\textwidth]{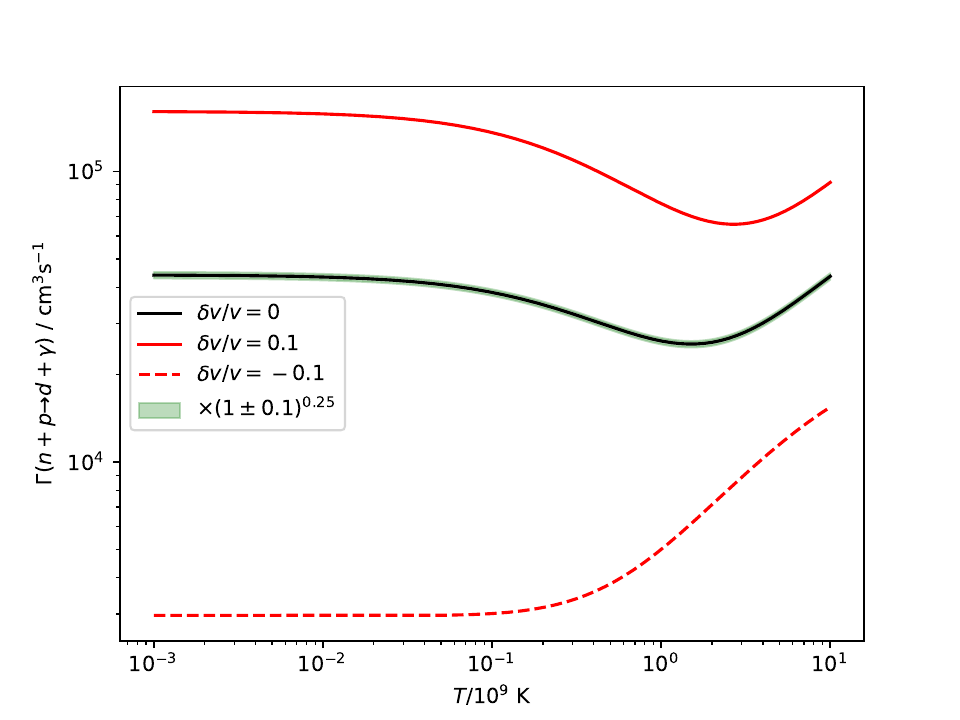}
    \caption{Reaction rate for $n + p \to d + \gamma$ for $\delta v / v = 0, \pm 0.1$ calculated from pionless EFT \cite{Rupak-2000} with nuclear parameters changed as described in the text. The shaded green band gives the original rate mutliplied by $\left(1 \pm 0.1 \right)^{0.25}$, so the range in which the  $n + p \to d + \gamma$ rate varies in the calculation of~\cite{burns2024constraints}.}
    \label{fig:npdg}
\end{figure}

Finally, we can use the pionless effective field theory (EFT) approach by \cite{Rupak-2000} for calculating the $n + p \to d + \gamma$ reaction rate, which is the first nuclear reaction taking place in BBN and therefore quite important. In \cite{burns2024constraints}, it is assumed that all nuclear reaction rates scale like $\Lambda_\mathrm{QCD}$, so a factor of  $(1 + \delta v / v)^{0.25}$ is multiplied to every rate. While this might be a reasonable approximation, one can make use of the analytic expression for the $n+p\to d+\gamma$ rate from \cite{Rupak-2000} and plug in the nucleon mass, the ${}^1S_0$ scattering length, the deuteron binding energy and the effective range $\rho_d$ at a given value of the Higgs VEV. The deuteron effective range can be found from the binding energy and the triplet scattering length $a_t$ via:
\begin{equation}
    \rho_d = 2\frac{a_t \sqrt{B_d m_N} -1}{a_t m_N B_d}~.
\end{equation}
The effective range of the ${}^1S_0$ channel only comes in at higher orders and we can safely ignore changes in this parameter as well as possible scale effects in the isoscalar contributions to the cross section.
The resulting reaction rate for $\delta v / v = 0, \pm 0.1$ is shown in Fig.~\ref{fig:npdg}. Changing the Higgs VEV therefore has a huge impact on the $n + p \to d + \gamma$ reaction. We also see that the functional form is not consistent
with a simple scaling with $(1 + \delta v / v)^{0.25}$.

Before presenting our final results, let us summarize the differences in our calculation to the one in~\cite{burns2024constraints} and in \cite{Berengut:2013nh}. First, we use an updated
value for  the neutron-proton mass difference, as given by Eq.~\eqref{eq:mnp}. Second,
the quark mass dependence of the deuteron binding energy is obtained from lattice
QCD data with the help of the nucleon-nucleon scattering LETs~\cite{Baru:2015ira,Baru:2016evv}, whereas in~\cite{burns2024constraints} a one-boson exchange
model was used with the appropriate scalings of the various meson exchanges. Interestingly, the resulting Higgs VEV dependence of the deuteron binding energy comes out very similar in both approaches. We also used lattice QCD data to model the pion-mass-dependence of the nucleon mass and the axial-vector current, which was not taken into account in \cite{burns2024constraints}. In \cite{Berengut:2013nh}, the pion mass dependence of these quantities was derived from chiral perturbation theory and the deuteron binding energy was found using 
chiral nuclear EFT at leading order, given by the one-pion-exchange potential together with the short-range four-nucleon contact interactions without derivatives, the latter exhibiting sizeable uncertainties. 
Then, we normalized the weak $n \leftrightarrow p$ rates using the appropriately varied neutron lifetime including the change in the Fermi-integral, see Eqs.~\eqref{eq:tau_n} and \eqref{eq:Fermi-int}. The main difference to \cite{Berengut:2013nh} is that  only the linear dependence of the weak rates on the neutron lifetime was considered. This does not include temperature-dependent effects. Finally, we made use of the pionless EFT approach by \cite{Rupak-2000} to include effects of a varying Higgs VEV on the $n + p \to d + \gamma$ rate, as explained in this section.

\section{Results}
\label{sec:res}

\begin{figure}
    \centering
    \includegraphics[width = 0.49\textwidth]{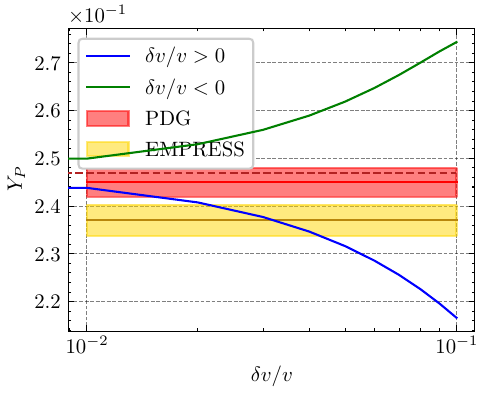}
    \includegraphics[width = 0.49\textwidth]{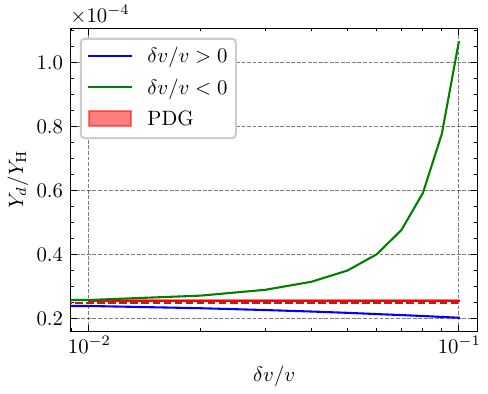}
    \caption{$^4$He (left) and $d$ (right)  abundance as a function of $\delta v / v$.}
    \label{fig:results}
\end{figure}

Finally, putting everything together, our results can be found in Fig.~\ref{fig:results}. What is most noticeable is the behaviour of the deuterium abundance: it increases drastically for negative values of $\delta v / v$, which is a result of the large changes in the $n + p \to d + \gamma$ rate. We expect the variation of the $^4$He abundance to be less pronounced than in~\cite{burns2024constraints} due to the different treatment of $Q_N$, see Sect.~\ref{sec:mnp}.
 Note that especially the deuterium abundance is sensitive to the choice of the rates for the 12 key reactions: in \texttt{PRyMordial}, one can choose between reaction rates collected by \texttt{PRIMAT} \cite{PRIMAT-companion} or by the \texttt{NACRE\,II} collaboration. As the former was chosen in \cite{burns2024constraints}\footnote{We are thankful to Anne-Kathrine Burns for supplying us with this information.}, we made the same choice so as to make the results more comparable. Using another set of rates would result in a shift of the abundances and alter the constraints for $\delta v / v$ accordingly. We did not use the improved rates from Ref.~\cite{Meissner:2023voo} since we are presently in the process of updating them and supplying theoretical uncertainties. This would go beyond the scope of this investigation.

The improved constraints for the Higgs VEV variation that can be derived by our calculation are collected in Tab.~\ref{tab:constraints}. We obtain these constraints by comparing our results to the observed values for the $^4$He and $d$ abundances given by the PDG \cite{PDG_Obs}. For $^4$He we also present constraints coming from the recently published values by the EMPRESS collaboration \cite{Matsumoto_2022}, which was also done in \cite{burns2024constraints}. The drastic increase of the deuterium abundance for negative $\delta v / v$ results in a very narrow range of possible variations of the Higgs VEV. This range is, however, included in the interval of possible $\delta v / v$ that can explain the PDG $^4$He abundance within $2\sigma$. 
We find for the linear dependence of the abundances on the Higgs VEV
\begin{equation}
    \dv{\ln Y_d}{\ln v} = -11.17\qc \dv{\ln Y_P}{\ln v} = -1.20.
\end{equation} 
We do note that the deuterium abundance can, of course, not be described by a linear function in $\delta v$, as is evident from Fig.\,\ref{fig:results}. In Ref.~\cite{Berengut:2013nh} the values of $\num{1.9\pm3.4}$ and $\num{-4.0\pm0.3}$ were found for deuterium and $^4$He, respectively. This discrepancy comes mainly from the treatment of the $n+p\to d+\gamma$ rate and from including temperature-dependent effects in the $n\leftrightarrow p$ rate. The dependences of the deuteron binding energy and the $^1S_0$ scattering length are also very different.
Note, however, that our treatment goes beyond the linear approximation and also, we have made clear improvements in the calculations
of the Higgs VEV dependence of a number of pertinent quantities. Also, our
bounds are dominated by the deuteron abundance and are more stringent than
found before.
\begin{table}
    \centering
    \begin{tabular}{|l|c|c|}\hline
         &  $^4$He & $d$\\ \hline
         PDG & $ \delta v / v \in [ -0.0140, 0.0264]$ & $\delta v / v \in [ -0.0047, -0.0013]$ \\
         EMPRESS & $\delta v / v \in [0.0111, 0.0546]$ & \\ \hline
         PDG & $ \delta v / v \in [ -0.0128, 0.0247]$ & $ \delta v / v \in [ -0.0592, -0.0180]$ \\
         EMPRESS & $ \delta v / v \in [0.0103, 0.0521]$ & \\ \hline 
    \end{tabular}
    \caption{Upper panel: Constraints on the Higgs VEV variation $\delta v / v$ derived from $2\sigma$ bounds of the observed values given in the PDG \cite{PDG_Obs} and by the EMPRESS collaboration \cite{Matsumoto_2022} for the abundances of $^4$He and deuterium. Lower panel: Constraints obtained from a calculation where  the $n + p \to d + \gamma$ rate is treated like in \cite{burns2024constraints}, see Fig.~\ref{fig:results_WOnpdg}.}
    \label{tab:constraints}
\end{table}

\begin{figure}
    \centering
    \includegraphics[width = 0.49\textwidth]{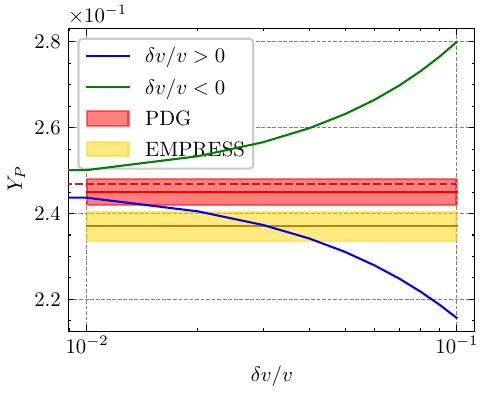}
    \includegraphics[width = 0.49\textwidth]{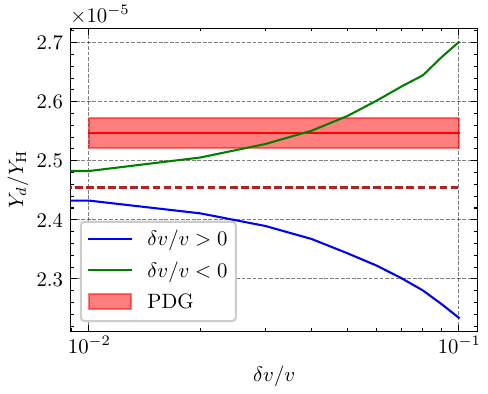}
    \caption{$^4$He (left) and $d$ (right) abundance as a function of $\delta v / v$ not including changes in the $n+p\to d +\gamma$ rate as would be implied by pionless EFT but scaling it like $\Lambda_\mathrm{QCD}$.}
    \label{fig:results_WOnpdg}
\end{figure}

In Fig.~\ref{fig:results_WOnpdg} we show the evolution of the abundances for different values of the Higgs VEV not including effects from the $n + p \to d + \gamma$ rate as would be implied by \cite{Rupak-2000} but scaling it like all other rates proportional to the QCD scale $\Lambda_\mathrm{QCD}$, i.e, as $ \sim (1+\delta v/v)^{0.25}$. The result for the deuterium abundance is markedly different from the more realistic calculation and now the constraints for $^4$He and $d$ contradict each other.

For the constraints coming from the EMPRESS data, we can confirm the result of \cite{burns2024constraints} that tuning the Higgs VEV so that the predicted $^4$He abundance matches the new EMPRESS result would worsen the deviation between theory and experiment for the deuterium abundance significantly. In the future,
we will use the methodology developed in Refs.~\cite{Meissner:2022dhg,Meissner:2023voo} and here to upgrade the constraints on combined 
variations of the Higgs VEV and the electromagnetic fine-structure constant $\alpha$ from the BBN abundances of all light elements. Also,
a fresh look on possible variations of the Yukawa couplings in the spirit of~\cite{Coc:2006sx} would be an interesting venue.

\section*{Acknowledgements}
We would like to thank Bernard Metsch for helpful discussions. 
We also thank Anne-Katherine Burns and  Marc Sher for clarifications on their work.
This work was supported in part by the European
Research Council (ERC) under the European Union's Horizon 2020 research
and innovation programme (grant agreement No. 101018170),
by DFG and NSFC through funds provided to the
Sino-German CRC 110 ``Symmetries and the Emergence of Structure in QCD" (NSFC
Grant No.~11621131001, Project-ID 196253076 - TRR 110).
The work of UGM was supported in part by the CAS President's International
Fellowship Initiative (PIFI) (Grant No.~2018DM0034).

\appendix

\section{Deuteron binding energy from a one-boson-exchange potential}
\label{sec:A}
\begin{figure}
    \centering
    \includegraphics{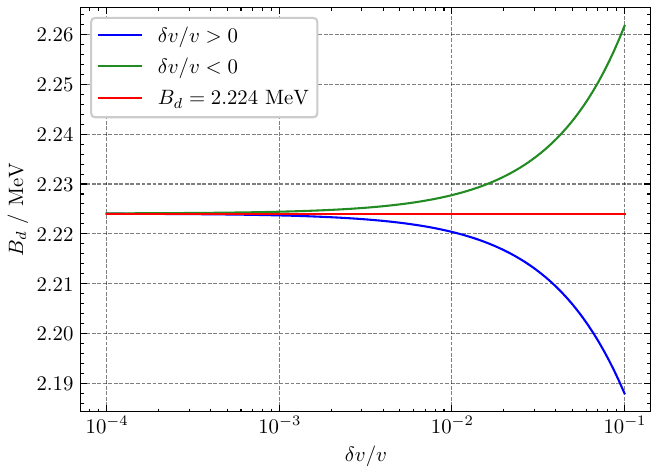}
    \caption{Deuteron binding energy as a function of $\delta v / v$ calculated using the OBEP from \cite{Lee:2020tmi}.}
    \label{fig:Bd_mesonexchange}
\end{figure}

In Sect.~\ref{sec:deut} we used lattice QCD data to model the pion-mass dependence of the deuteron binding energy which
was similar to the dependence found in~\cite{burns2024constraints} from calculating the binding energy using a one-boson exchange
potential (OBEP).  The authors of~\cite{burns2024constraints}  use a superposition of three Yukawa functions corresponding to 
pion-, sigma- and omega-meson exchanges, where the first two are attractive and the last one supplies the required repulsion. The mass of the $\sigma$ is varied so as to get the appropriate $B_d$ at the physical pion mass. Note that no meson-nucleon form factors are employed in that calculation.  We try to confirm their results by implementing the OBEP potential described in \cite{Lee:2020tmi}, where also contributions
from $\rho$-meson exchange are considered and a common form factor for all meson-nucleon vertices is utilizied. 
For more details, we refer to the textbook~\cite{Ericson:1988gk} (and references therein).
In \cite{Lee:2020tmi},  the small contribution from the $D$-wave for the ${}^3S_1$-channel is neglected, that is one works strictly with
angular momentum $L = 0$. The potentials corresponding to the different meson exchanges are then given in coordinate space as:
\begin{align}
    V_\pi(r) &= (\boldsymbol{\tau_1}\cdot\boldsymbol{\tau_2}) (\boldsymbol{\sigma_1}\cdot\boldsymbol{\sigma_2}) \frac{g^2_{\pi NN}}{4\pi} \frac{1}{12} \left( \frac{M_\pi}{m_N}\right)^2 \frac{e^{-M_\pi r}}{r}~, \\
    V_\sigma(r) &= - \frac{g^2_{\sigma NN}}{4\pi} \left( 1 - \frac{1}{4}\left( \frac{M_\sigma}{m_N}\right)^2\right) \frac{e^{-M_\sigma r}}{r}~, \\
    V_\omega(r) &= \frac{g^2_{\omega NN}}{4\pi} \left( 1 + \frac{1}{2} \left( \frac{M_\omega}{m_N}\right)^2 \left[ 1 + \frac{1}{3}(\boldsymbol{\sigma_1}\cdot\boldsymbol{\sigma_2})\right]\right) \frac{e^{-M_\omega r}}{r}~, \\
    V_\rho(r) &= (\boldsymbol{\tau_1}\cdot\boldsymbol{\tau_2}) \frac{g^2_{\rho NN}}{4\pi} \left( 1 + \frac{1}{2} \left( \frac{M_\rho}{m_N}\right)^2 \left[ 1 + \frac{g_T^\rho}{g_{\rho NN}} + \frac{1}{3} \left(1 + \frac{g_T^\rho}{g_{\rho NN}} \right)^2 (\boldsymbol{\sigma_1}\cdot\boldsymbol{\sigma_2})\right]\right) \frac{e^{-M_\rho r}}{r}~.
\end{align}
Here, $m_N$ is the average nucleon mass. The potential depends on the total spin $S$ and total isospin $I$ of the channel
through the factors $(\boldsymbol{\sigma_1}\cdot\boldsymbol{\sigma_2}) = 2S(S+1) -3$ and $(\boldsymbol{\tau_1}\cdot\boldsymbol{\tau_2}) = 2I(I+1) - 3$. 
The full potential is then
\begin{equation}
    V_\mathrm{OBE}(r) = \sum_{\alpha = \{\pi,\sigma,\omega,\rho\}} V_\alpha(r) + \frac{1}{4\pi} \frac{e^{-\Lambda r}}{r},
\end{equation}
where the cutoff $\Lambda$ regularizes the UV-divergence. In general, one takes different cut-offs for different mesons, but since
we are only interested in the deuteron binding energy, this simplified form suffices. The physical values for the meson mass are
\begin{equation}
  M_\pi = \SI{139.57}{\mega\electronvolt}\qc M_\sigma = \SI{550}{\mega\electronvolt}\qc M_\omega = \SI{783}{\mega\electronvolt}\qc
  M_\rho = \SI{769}{\mega\electronvolt},
\end{equation}
and they are scaled with the Higgs VEV as explained in \cite{burns2024constraints}: the pion mass with $(\delta v/v)^{1.25/2}$ and
the other masses as $(\delta v / v)^{0.25}$. 
The coupling constant $g_{\pi NN}$ is taken from the Goldberger-Treiman discrepancy~\cite{Fettes:1998ud}
\begin{equation}\label{GTR}
  g_{\pi NN}(\delta v) = \frac{g_A(\delta v) m_N(\delta v)}{F_\pi(\delta v)} \left(1 - \frac{2M_\pi^2(\delta v) \Bar{d}_{18}}{g_A(\delta v)}\right)\qc \Bar{d}_{18} = \SI{-0.47}{\per\giga\electronvolt\squared},
\end{equation}
where $m_N$ are $g_A$ given in Eqs.~\eqref{eq:mNMpi} and \eqref{eq:gAMpi} as a function of the pion mass. $F_\pi(M_\pi)$ can also be found from a fit to lattice data by \cite{PhysRevD.90.114504}:
\begin{equation}
    F_\pi(M_\pi) = F_\pi^\mathrm{phys} + \SI{0.138\pm 0.009}{\per\giga\electronvolt} (M_\pi^2-(M_\pi^\mathrm{phys})^2) - \SI{0.068\pm 0.043}{\per\cubic\giga\electronvolt} (M_\pi^2-(M_\pi^\mathrm{phys})^2)^2,
\end{equation}
with $M_\pi$ given in $\si{\giga\electronvolt}$.
Finally, we use the coupling constants given in \cite{Lee:2020tmi} as parameter set I:
\begin{equation}
    \frac{g^2_{\sigma NN}}{4\pi} = 14.17\qc \frac{g^2_{\omega NN}}{4\pi} = 20.0\qc \frac{g^2_{\rho NN}}{4\pi} = 0.80\qc \Lambda = \SI{1364}{\mega\electronvolt}
\end{equation}
and $g_T^\rho / g_{\rho NN} = 6.1$, and the tiny $\omega$ tensor coupling is neglected. We note that this model just describes the deuteron binding energy
and not any more the full S-wave phase shifts. For that, one would have to vary the cut-off for each meson exchange. However, for the calculation performed here such
a simplified treatment is sufficient.
The resulting change in the deuteron binding energy, when we vary the Higgs VEV by $\pm 10\%$, is much smaller than what was given in \cite{burns2024constraints} and also has the opposite sign, see Fig.~\ref{fig:Bd_mesonexchange}. Note further that keeping $g_{\pi NN}$ constant instead
of varying it as in Eq.\eqref{GTR} has only a mild influence on the results.

\begin{figure}
    \centering
    \includegraphics[width = 0.8\textwidth]{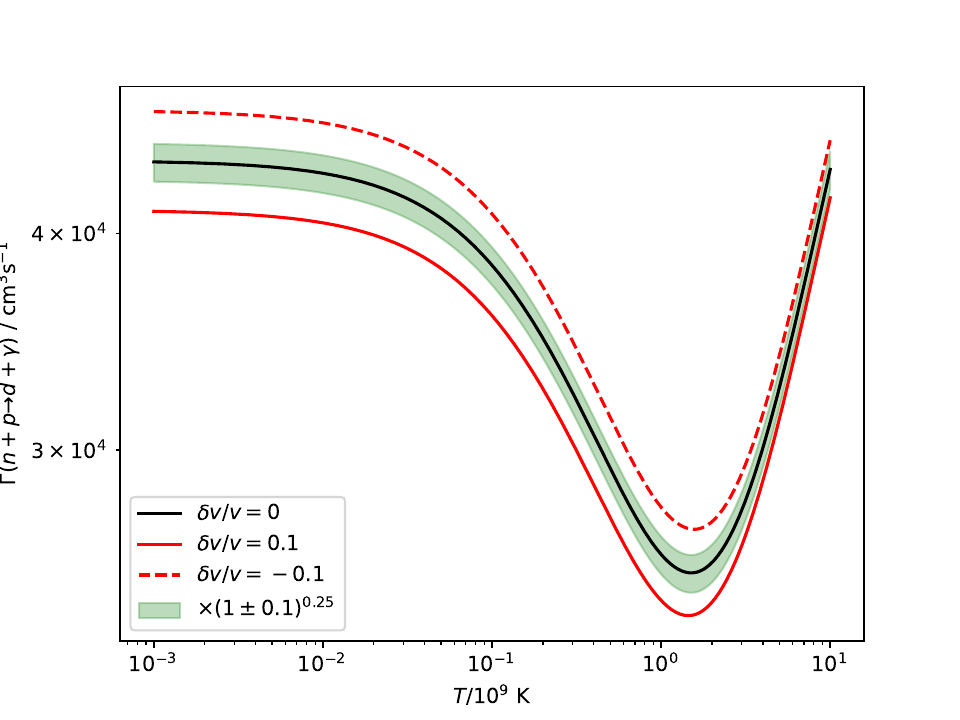}
    \caption{Reaction rate for $n + p \to d + \gamma$ for $\delta v / v = 0, \pm 0.1$ calculated from pionless effective field
      theory \cite{Rupak-2000} with nuclear parameters changed as described in the text, but the binding energy taken from a
      OBEP calculation. The shaded green band gives the original rate mutliplied by $\left(1 \pm 0.1 \right)^{0.25}$,
      so the range in which the  $n + p \to d + \gamma$ rate varies in the calculation of \cite{burns2024constraints}.}
    \label{fig:npdg_Meson}
\end{figure}

Plugging in this deuteron binding energy dependence, we find that the $n+p\to d+\gamma$ rate does not vary as much as before
and resembles more the $(1 + \delta v/v)^{0.25}$-scaling that was used in \cite{burns2024constraints}. This is presented
in Fig.~\ref{fig:npdg_Meson}.

\begin{figure}
    \centering
    \includegraphics[width = 0.49\textwidth]{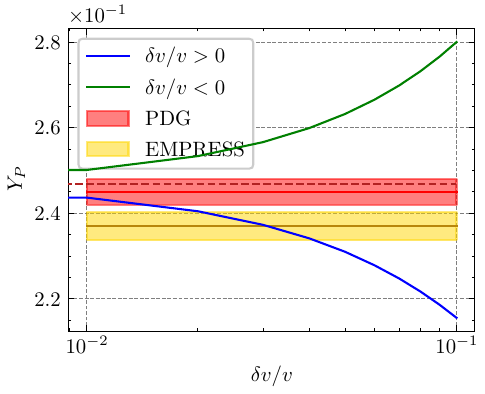}
    \includegraphics[width = 0.49\textwidth]{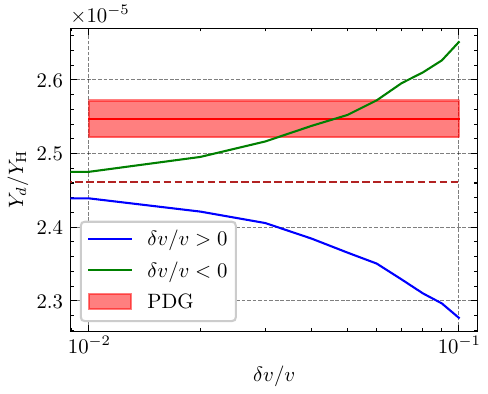}
    \caption{$^4$He (left) and deuterium (right) abundance as a function of $\delta v / v$ for the deuteron binding
      energy calculated using an OBEP.}
    \label{fig:results_Meson}
\end{figure}

The resulting $^4$He and $d$ abundances for this calculation can the found in Fig.~\ref{fig:results_Meson}. They are close to
the results obtained by simply scaling the $n+p\to d +\gamma$ rate as described in \cite{burns2024constraints}, which is
evident when comparing Fig.~\ref{fig:results_Meson} to Fig.~\ref{fig:results_WOnpdg}.

\section{Comparison to the literature}
\label{sec:B}
In this section we presents in detail how our work differs from and improves upon earlier publications. 

Dixit and Sher (1988) \cite{Dixit:1987at} and Scherrer and Spergel (1993) \cite{Scherrer:1992na} were one of the firsts to study effects of a possible Higgs VEV dependence on primordial abundances in the framework of BBN. In both papers (and all that we are mentioning in this section), the calculation for the neutron-proton mass difference uses the values from~\cite{Gasser:1982ap} for the electromagnetic and strong contribution, while we use the more recent ones from~\cite{QnAlpha}. It is not clear to us if the authors included temperature-dependent effects in the weak $n\leftrightarrow p$ rates and effects from a changing deuteron binding energy or other nucleon-nucleon scattering parameters on the $n + p \to d + \gamma$ rate, as well as a varying nucleon mass, were not considered. In \cite{Dixit:1987at}, the pion mass dependence on the Higgs VEV does not include any variation of the strong scale and for the deuteron binding energy dependence on the pion mass the authors made a simple estimation. In \cite{Scherrer:1992na} a changing pion mass or deuteron binding energy was not mentioned at all. 

Yoo and Scherrer \cite{Yoo:2002vw} then improved the treatment of the deuteron binding energy in 2003, assuming a linear dependence on the pion mass. The authors also derived constraints on the variation of the Higgs VEV from CMB measurements in addition to BBN simulations. Li and Chu (2006) \cite{PhysRevD.73.023509} employed much of the methodology from \cite{Yoo:2002vw}, but included now the variation of the deuteron binding energy (but not other scattering parameters) in a pionless EFT approach for the $n + p \to d + \gamma$ rate.

Campbell and Olive studied the dependence of primordial abundances on the VEV of some dilation field in 1994, including effects of a varying Higgs VEV \cite{Campbell:1994bf}. In our understanding they did not, however, implement their methods into a BBN code but used a simple approximated expression for the Helium-4 abundance. Similarly, Dmitriev and Flambaum (2003) \cite{PhysRevD.67.063513} and Flambaum and Shuryak (2003) \cite{Flambaum:2002de} considered only changes in the quark masses and the strong scale and do not use a BBN code but find constraints for example from the Helium-5 binding. Dmitriev, Flambaum and Webb (2004) \cite{PhysRevD.69.063506} consider only the change in the deuteron binding energy and use the simple scaling
    \begin{equation}
        \expval{\sigma(n+p\to d+\gamma) v} \propto B_d^{5/2}
    \end{equation}
for the $n + p \to d + \gamma$ rate. Although Bedaque, Luu and Platter \cite{Bedaque:2010hr} found this scaling to be rather accurate for the deuteron binding energy, it does not include effects from changing other nucleon-nucleon scattering paramters. 

A Higgs VEV variation was also part of the work of Coc et al. in 2006 \cite{Coc:2006sx}. The employ a scaling of the $n + p \to d + \gamma$ rate like in \cite{PhysRevD.69.063506} and consider a linear dependence of the deuteron binding energy either on the $\sigma$- and $\omega$-meson mass and the nucleon mass or the same linear dependence on the pion mass as in \cite{Yoo:2002vw}. These authors considerered
combined variations of the Higgs VEV and the Yukawa couplings. Using some approximate scaling relations, as
they appear in the discussed unified theories, limits on the variations of the Yukawa couplings are displayed. Thus, a direct comparison with our work is not possible.

In their extensive work on the variation of fundamental constants in the BBN framework, Dent, Stern and Wetterich \cite{Dent:2007zu}, carried out the Higgs VEV variation independently of varying the fermion masses. One would have to adjust the response matrices accordingly to properly compare our results. The authors otherwise used a very similar approach to \cite{Yoo:2002vw}. 

Finally, the differences to the works of Bedaque, Luu and Platter \cite{Bedaque:2010hr}, Berengut et al. \cite{Berengut:2013nh} and, of course, to the newest publication of Burns et al.~\cite{burns2024constraints}
were explained in detail in the text. 

The ways we have improved the methodology of the publications listed here is summarised at the end of section \ref{sec:npdg}.

\bibliographystyle{JHEP} 
\bibliography{refs}%

\providecommand{\href}[2]{#2}\begingroup\raggedright\begin{thebibliography}{10}

\bibitem{burns2024constraints}
A.-K. Burns, V.~Keus, M.~Sher, and T.~M.~P. Tait, {\it Constraints on variation
  of the weak scale from big bang nucleosynthesis},  2024.

\bibitem{Hogan_2000}
C.~J. Hogan, {\it Why the universe is just so},  {\em Reviews of Modern
  Physics} {\bf 72} (Oct., 2000) 1149–1161.

\bibitem{Uzan_2003}
J.-P. Uzan, {\it The fundamental constants and their variation: observational
  and theoretical status},  {\em Reviews of Modern Physics} {\bf 75} (Apr.,
  2003) 403–455.

\bibitem{Schellekens_2013}
A.~N. Schellekens, {\it Life at the interface of particle physics and string
  theory},  {\em Reviews of Modern Physics} {\bf 85} (Oct., 2013) 1491–1540.

\bibitem{Meissner_2015}
U.-G. Meißner, {\it Anthropic considerations in nuclear physics},  {\em
  Science Bulletin} {\bf 60} (Jan., 2015) 43–54.

\bibitem{Donoghue_2016}
J.~F. Donoghue, {\it The multiverse and particle physics},  {\em Annual Review
  of Nuclear and Particle Science} {\bf 66} (Oct., 2016) 1–21.

\bibitem{Adams_2019}
F.~C. Adams, {\it The degree of fine-tuning in our universe — and others},
  {\em Physics Reports} {\bf 807} (May, 2019) 1–111.

\bibitem{burns2023prymordial}
A.-K. Burns, T.~M.~P. Tait, and M.~Valli, {\it Prymordial: The first three
  minutes, within and beyond the standard model},  2023.

\bibitem{Matsumoto_2022}
A.~Matsumoto, M.~Ouchi, K.~Nakajima, M.~Kawasaki, K.~Murai, K.~Motohara,
  Y.~Harikane, Y.~Ono, K.~Kushibiki, S.~Koyama, S.~Aoyama, M.~Konishi,
  H.~Takahashi, Y.~Isobe, H.~Umeda, Y.~Sugahara, M.~Onodera, K.~Nagamine,
  H.~Kusakabe, Y.~Hirai, T.~J. Moriya, T.~Shibuya, Y.~Komiyama, K.~Fukushima,
  S.~Fujimoto, T.~Hattori, K.~Hayashi, A.~K. Inoue, S.~Kikuchihara, T.~Kojima,
  Y.~Koyama, C.-H. Lee, K.~Mawatari, T.~Miyata, T.~Nagao, S.~Ozaki, M.~Rauch,
  T.~Saito, A.~Suzuki, T.~T. Takeuchi, M.~Umemura, Y.~Xu, K.~Yabe, Y.~Zhang,
  and Y.~Yoshii, {\it Empress. viii. a new determination of primordial he
  abundance with extremely metal-poor galaxies: A suggestion of the lepton
  asymmetry and implications for the hubble tension},  {\em The Astrophysical
  Journal} {\bf 941} (Dec., 2022) 167.

\bibitem{PDG_Obs}
R.~L. Workman et~al., {\it {Review of Particle Physics}},  {\em PTEP} (08,
  2022)
  [\href{http://arxiv.org/abs/https://academic.oup.com/ptep/article-pdf/2022/8/083C01/49175539/ptac097.pdf}{{\tt
  https://academic.oup.com/ptep/article-pdf/2022/8/083C01/49175539/ptac097.pdf}}].
  083C01.

\bibitem{Dixit:1987at}
V.~V. Dixit and M.~Sher, {\it {Variation of the Fermi Constant and Primordial
  Nucleosynthesis}},  {\em Phys. Rev. D} {\bf 37} (1988) 1097.

\bibitem{Scherrer:1992na}
R.~J. Scherrer and D.~N. Spergel, {\it {How constant is the Fermi coupling
  constant?}},  {\em Phys. Rev. D} {\bf 47} (1993) 4774--4777.

\bibitem{Yoo:2002vw}
J.~Yoo and R.~J. Scherrer, {\it {Big bang nucleosynthesis and cosmic microwave
  background constraints on the time variation of the Higgs vacuum expectation
  value}},  {\em Phys. Rev. D} {\bf 67} (2003) 043517,
  [\href{http://arxiv.org/abs/astro-ph/0211545}{{\tt astro-ph/0211545}}].

\bibitem{PhysRevD.73.023509}
B.~Li and M.~C. Chu, {\it Big-bang nucleosynthesis with an evolving radion in
  the brane world scenario},  {\em Phys. Rev. D} {\bf 73} (Jan, 2006) 023509.

\bibitem{Dent:2007zu}
T.~Dent, S.~Stern, and C.~Wetterich, {\it {Primordial nucleosynthesis as a
  probe of fundamental physics parameters}},  {\em Phys. Rev. D} {\bf 76}
  (2007) 063513, [\href{http://arxiv.org/abs/0705.0696}{{\tt
  arXiv:0705.0696}}].

\bibitem{Bedaque:2010hr}
P.~F. Bedaque, T.~Luu, and L.~Platter, {\it {Quark mass variation constraints
  from Big Bang nucleosynthesis}},  {\em Phys. Rev. C} {\bf 83} (2011) 045803,
  [\href{http://arxiv.org/abs/1012.3840}{{\tt arXiv:1012.3840}}].

\bibitem{Berengut:2013nh}
J.~C. Berengut, E.~Epelbaum, V.~V. Flambaum, C.~Hanhart, U.~G. Mei{\ss}ner,
  J.~Nebreda, and J.~R. Pelaez, {\it {Varying the light quark mass: impact on
  the nuclear force and Big Bang nucleosynthesis}},  {\em Phys. Rev. D} {\bf
  87} (2013), no.~8 085018, [\href{http://arxiv.org/abs/1301.1738}{{\tt
  arXiv:1301.1738}}].

\bibitem{Beane:2002vs}
S.~R. Beane and M.~J. Savage, {\it {Variation of fundamental couplings and
  nuclear forces}},  {\em Nucl. Phys. A} {\bf 713} (2003) 148--164,
  [\href{http://arxiv.org/abs/hep-ph/0206113}{{\tt hep-ph/0206113}}].

\bibitem{Epelbaum:2002gb}
E.~Epelbaum, U.-G. Mei{\ss}ner, and W.~Gloeckle, {\it {Nuclear forces in the
  chiral limit}},  {\em Nucl. Phys. A} {\bf 714} (2003) 535--574,
  [\href{http://arxiv.org/abs/nucl-th/0207089}{{\tt nucl-th/0207089}}].

\bibitem{Epelbaum:2001fm}
E.~Epelbaum, U.~G. Mei{\ss}ner, W.~Gloeckle, and C.~Elster, {\it {Resonance
  saturation for four nucleon operators}},  {\em Phys. Rev. C} {\bf 65} (2002)
  044001, [\href{http://arxiv.org/abs/nucl-th/0106007}{{\tt nucl-th/0106007}}].

\bibitem{Baru:2015ira}
V.~Baru, E.~Epelbaum, A.~A. Filin, and J.~Gegelia, {\it {Low-energy theorems
  for nucleon-nucleon scattering at unphysical pion masses}},  {\em Phys. Rev.
  C} {\bf 92} (2015), no.~1 014001,
  [\href{http://arxiv.org/abs/1504.07852}{{\tt arXiv:1504.07852}}].

\bibitem{Baru:2016evv}
V.~Baru, E.~Epelbaum, and A.~A. Filin, {\it {Low-energy theorems for
  nucleon-nucleon scattering at $M_\pi=450$ MeV}},  {\em Phys. Rev. C} {\bf 94}
  (2016), no.~1 014001, [\href{http://arxiv.org/abs/1604.02551}{{\tt
  arXiv:1604.02551}}].

\bibitem{Gasser:1982ap}
J.~Gasser and H.~Leutwyler, {\it {Quark Masses}},  {\em Phys. Rept.} {\bf 87}
  (1982) 77--169.

\bibitem{QnAlpha}
J.~Gasser, H.~Leutwyler, and A.~Rusetsky, {\it {On the mass difference between
  proton and neutron}},  {\em Phys. Lett. B} {\bf 814} (2021) 136087,
  [\href{http://arxiv.org/abs/2003.13612}{{\tt arXiv:2003.13612}}].

\bibitem{Meissner:2022cbi}
U.-G. Mei\ss{}ner and A.~Rusetsky, {\em {Effective Field Theories}}.
\newblock Cambridge University Press, 8, 2022.

\bibitem{Walker-Loud:2014iea}
A.~Walker-Loud, {\it {Nuclear Physics Review}},  {\em PoS} {\bf LATTICE2013}
  (2014) 013, [\href{http://arxiv.org/abs/1401.8259}{{\tt arXiv:1401.8259}}].

\bibitem{Bernard:2006te}
V.~Bernard and U.-G. Mei{\ss}ner, {\it {The Nucleon axial-vector coupling
  beyond one loop}},  {\em Phys. Lett. B} {\bf 639} (2006) 278--282,
  [\href{http://arxiv.org/abs/hep-lat/0605010}{{\tt hep-lat/0605010}}].

\bibitem{Alexandrou:2013jsa}
C.~Alexandrou, M.~Constantinou, V.~Drach, K.~Jansen, C.~Kallidonis, and
  G.~Koutsou, {\it {Nucleon generalized form factors with twisted mass
  fermions}},  {\em PoS} {\bf LATTICE2013} (2014) 292,
  [\href{http://arxiv.org/abs/1312.2874}{{\tt arXiv:1312.2874}}].

\bibitem{Alexandrou:2014wca}
C.~Alexandrou, M.~Constantinou, K.~Hadjiyiannakou, K.~Jansen, C.~Kallidonis,
  and G.~Koutsou, {\it {Nucleon observables and axial charges of other baryons
  using twisted mass fermions}},  {\em PoS} {\bf LATTICE2014} (2015) 151,
  [\href{http://arxiv.org/abs/1411.3494}{{\tt arXiv:1411.3494}}].

\bibitem{Chang:2018uxx}
C.~C. Chang et~al., {\it {A per-cent-level determination of the nucleon axial
  coupling from quantum chromodynamics}},  {\em Nature} {\bf 558} (2018),
  no.~7708 91--94, [\href{http://arxiv.org/abs/1805.12130}{{\tt
  arXiv:1805.12130}}].

\bibitem{MEISSNER1988213}
U.-G. Mei{\ss}ner, {\it Low-energy hadron physics from effective chiral
  lagrangians with vector mesons},  {\em Physics Reports} {\bf 161} (1988),
  no.~5 213--361.

\bibitem{Hanhart:2008mx}
C.~Hanhart, J.~R. Pelaez, and G.~Rios, {\it {Quark mass dependence of the rho
  and sigma from dispersion relations and Chiral Perturbation Theory}},  {\em
  Phys. Rev. Lett.} {\bf 100} (2008) 152001,
  [\href{http://arxiv.org/abs/0801.2871}{{\tt arXiv:0801.2871}}].

\bibitem{Lahde:2019yvr}
T.~A. L\"ahde, U.-G. Mei\ss{}ner, and E.~Epelbaum, {\it {An update on
  fine-tunings in the triple-alpha process}},  {\em Eur. Phys. J. A} {\bf 56}
  (2020), no.~3 89, [\href{http://arxiv.org/abs/1906.00607}{{\tt
  arXiv:1906.00607}}].

\bibitem{Wagman:2017tmp}
M.~L. Wagman, F.~Winter, E.~Chang, Z.~Davoudi, W.~Detmold, K.~Orginos, M.~J.
  Savage, and P.~E. Shanahan, {\it {Baryon-Baryon Interactions and Spin-Flavor
  Symmetry from Lattice Quantum Chromodynamics}},  {\em Phys. Rev. D} {\bf 96}
  (2017), no.~11 114510, [\href{http://arxiv.org/abs/1706.06550}{{\tt
  arXiv:1706.06550}}].

\bibitem{NPLQCD:2020lxg}
{\bf NPLQCD} Collaboration, M.~Illa et~al., {\it {Low-energy scattering and
  effective interactions of two baryons at $m_{\pi}\sim 450$ MeV from lattice
  quantum chromodynamics}},  {\em Phys. Rev. D} {\bf 103} (2021), no.~5 054508,
  [\href{http://arxiv.org/abs/2009.12357}{{\tt arXiv:2009.12357}}].

\bibitem{Amarasinghe:2021lqa}
S.~Amarasinghe, R.~Baghdadi, Z.~Davoudi, W.~Detmold, M.~Illa, A.~Parreno, A.~V.
  Pochinsky, P.~E. Shanahan, and M.~L. Wagman, {\it {Variational study of
  two-nucleon systems with lattice QCD}},  {\em Phys. Rev. D} {\bf 107} (2023),
  no.~9 094508, [\href{http://arxiv.org/abs/2108.10835}{{\tt
  arXiv:2108.10835}}].

\bibitem{Berkowitz:2015eaa}
E.~Berkowitz, T.~Kurth, A.~Nicholson, B.~Joo, E.~Rinaldi, M.~Strother, P.~M.
  Vranas, and A.~Walker-Loud, {\it {Two-Nucleon Higher Partial-Wave Scattering
  from Lattice QCD}},  {\em Phys. Lett. B} {\bf 765} (2017) 285--292,
  [\href{http://arxiv.org/abs/1508.00886}{{\tt arXiv:1508.00886}}].

\bibitem{NPLQCD:2011naw}
{\bf NPLQCD} Collaboration, S.~R. Beane, E.~Chang, W.~Detmold, H.~W. Lin, T.~C.
  Luu, K.~Orginos, A.~Parreno, M.~J. Savage, A.~Torok, and A.~Walker-Loud, {\it
  {The Deuteron and Exotic Two-Body Bound States from Lattice QCD}},  {\em
  Phys. Rev. D} {\bf 85} (2012) 054511,
  [\href{http://arxiv.org/abs/1109.2889}{{\tt arXiv:1109.2889}}].

\bibitem{Yamazaki:2015asa}
T.~Yamazaki, K.-i. Ishikawa, Y.~Kuramashi, and A.~Ukawa, {\it {Study of quark
  mass dependence of binding energy for light nuclei in 2+1 flavor lattice
  QCD}},  {\em Phys. Rev. D} {\bf 92} (2015), no.~1 014501,
  [\href{http://arxiv.org/abs/1502.04182}{{\tt arXiv:1502.04182}}].

\bibitem{Orginos:2015aya}
K.~Orginos, A.~Parreno, M.~J. Savage, S.~R. Beane, E.~Chang, and W.~Detmold,
  {\it {Two nucleon systems at $m_\pi\sim 450~{\rm MeV}$ from lattice QCD}},
  {\em Phys. Rev. D} {\bf 92} (2015), no.~11 114512,
  [\href{http://arxiv.org/abs/1508.07583}{{\tt arXiv:1508.07583}}]. [Erratum:
  Phys.Rev.D 102, 039903 (2020)].

\bibitem{NPLQCD:2013bqy}
{\bf NPLQCD} Collaboration, S.~R. Beane et~al., {\it {Nucleon-Nucleon
  Scattering Parameters in the Limit of SU(3) Flavor Symmetry}},  {\em Phys.
  Rev. C} {\bf 88} (2013), no.~2 024003,
  [\href{http://arxiv.org/abs/1301.5790}{{\tt arXiv:1301.5790}}].

\bibitem{Yamazaki:2012hi}
T.~Yamazaki, K.-i. Ishikawa, Y.~Kuramashi, and A.~Ukawa, {\it {Helium nuclei,
  deuteron and dineutron in 2+1 flavor lattice QCD}},  {\em Phys. Rev. D} {\bf
  86} (2012) 074514, [\href{http://arxiv.org/abs/1207.4277}{{\tt
  arXiv:1207.4277}}].

\bibitem{Segre-gamow}
E.~Segr{\`e}, {\em Nuclei and Particles}, vol.~2.
\newblock Basic Books, 1964.

\bibitem{Meissner:2023voo}
U.-G. Mei\ss{}ner, B.~C. Metsch, and H.~Meyer, {\it {The electromagnetic
  fine-structure constant in primordial nucleosynthesis revisited}},  {\em Eur.
  Phys. J. A} {\bf 59} (2023), no.~10 223,
  [\href{http://arxiv.org/abs/2305.15849}{{\tt arXiv:2305.15849}}].

\bibitem{Rupak-2000}
G.~Rupak, {\it Precision calculation of $np \to d\gamma$ cross section for
  big-bang nucleosynthesis},  {\em Nucl. Phys. A} {\bf 678} (2000), no.~4
  405--423.

\bibitem{PRIMAT-companion}
C.~Pitrou, A.~Coc, J.-P. Uzan, and E.~Vangioni, {\it Precision big bang
  nucleosynthesis with improved helium-4 predictions},  {\em Phys. Rept.} {\bf
  754} (2018) 1--66.

\bibitem{Meissner:2022dhg}
U.-G. Mei\ss{}ner and B.~C. Metsch, {\it {Probing nuclear observables via
  primordial nucleosynthesis}},  {\em Eur. Phys. J. A} {\bf 58} (2022), no.~11
  212, [\href{http://arxiv.org/abs/2208.12600}{{\tt arXiv:2208.12600}}].

\bibitem{Coc:2006sx}
A.~Coc, N.~J. Nunes, K.~A. Olive, J.-P. Uzan, and E.~Vangioni, {\it {Coupled
  Variations of Fundamental Couplings and Primordial Nucleosynthesis}},  {\em
  Phys. Rev. D} {\bf 76} (2007) 023511,
  [\href{http://arxiv.org/abs/astro-ph/0610733}{{\tt astro-ph/0610733}}].

\bibitem{Lee:2020tmi}
D.~Lee, U.-G. Mei\ss{}ner, K.~A. Olive, M.~Shifman, and T.~Vonk, {\it
  {\ensuremath{\theta} -dependence of light nuclei and nucleosynthesis}},  {\em
  Phys. Rev. Res.} {\bf 2} (2020), no.~3 033392,
  [\href{http://arxiv.org/abs/2006.12321}{{\tt arXiv:2006.12321}}].

\bibitem{Ericson:1988gk}
T.~E.~O. Ericson and W.~Weise, {\em {Pions and Nuclei}}.
\newblock Clarendon Press, Oxford, UK, 1988.

\bibitem{Fettes:1998ud}
N.~Fettes, U.-G. Mei{\ss}ner, and S.~Steininger, {\it {Pion - nucleon
  scattering in chiral perturbation theory. 1. Isospin symmetric case}},  {\em
  Nucl. Phys. A} {\bf 640} (1998) 199--234,
  [\href{http://arxiv.org/abs/hep-ph/9803266}{{\tt hep-ph/9803266}}].

\bibitem{PhysRevD.90.114504}
{\bf Budapest-Marseille-Wuppertal Collaboration} Collaboration, S.~D\"urr,
  Z.~Fodor, C.~Hoelbling, S.~Krieg, T.~Kurth, L.~Lellouch, T.~Lippert,
  R.~Malak, T.~M\'etivet, A.~Portelli, A.~Sastre, and K.~Szab\'o, {\it Lattice
  qcd at the physical point meets $su(2)$ chiral perturbation theory},  {\em
  Phys. Rev. D} {\bf 90} (Dec, 2014) 114504.

\bibitem{Campbell:1994bf}
B.~A. Campbell and K.~A. Olive, {\it {Nucleosynthesis and the time dependence
  of fundamental couplings}},  {\em Phys. Lett. B} {\bf 345} (1995) 429--434,
  [\href{http://arxiv.org/abs/hep-ph/9411272}{{\tt hep-ph/9411272}}].

\bibitem{PhysRevD.67.063513}
V.~F. Dmitriev and V.~V. Flambaum, {\it Limits on cosmological variation of
  quark masses and strong interaction},  {\em Phys. Rev. D} {\bf 67} (Mar,
  2003) 063513.

\bibitem{Flambaum:2002de}
V.~V. Flambaum and E.~V. Shuryak, {\it {Limits on cosmological variation of
  strong interaction and quark masses from big bang nucleosynthesis, cosmic,
  laboratory and Oklo data}},  {\em Phys. Rev. D} {\bf 65} (2002) 103503,
  [\href{http://arxiv.org/abs/hep-ph/0201303}{{\tt hep-ph/0201303}}].

\bibitem{PhysRevD.69.063506}
V.~F. Dmitriev, V.~V. Flambaum, and J.~K. Webb, {\it Cosmological variation of
  the deuteron binding energy, strong interaction, and quark masses from big
  bang nucleosynthesis},  {\em Phys. Rev. D} {\bf 69} (Mar, 2004) 063506.

\end{thebibliography}\endgroup
\end{document}